\documentclass[12pt]{article}
\usepackage{amsmath,amssymb}
\usepackage{graphicx}
\makeatletter

    \@addtoreset{equation}{section}
\makeatother
\begin{document}

\thispagestyle{empty}

\vspace*{2cm}

\begin{center}
 {\LARGE {$\beta$-deformed matrix model\\[3mm] and Nekrasov partition function}}
\vskip2cm
{\large 
{Takahiro Nishinaka$^{1,}$\footnote{email: nishinak@post.kek.jp}
}
and\hspace{2mm} Chaiho Rim$^{2,}$\footnote{email: rimpine@sogang.ac.kr}
}
\vskip.5cm
{\it Department of Physics$^2$ and Center for Quantum Spacetime (CQUeST)$^{1,2}$
\\
Sogang University, Seoul 121-742, Korea}
\end{center}

\vskip2cm
\begin{abstract}
We study Penner type matrix models in relation with the Nekrasov partition function of  four dimensional $ \mathcal{N}=2, ~ SU(2)$ supersymmetric gauge theories with $N_F=2,3$ and $4$. By evaluating  the resolvent using the loop equation for general $\beta$, we explicitly construct the first half-genus correction to the free energy
and demonstrate the result coincides with the corresponding Nekrasov partition function with general $\Omega$-background, including higher instanton contributions
after modifying the relation of the Coulomb branch parameter with the filling fraction.
Our approach complements the proof using the Selberg integrals directly
which is useful to find the contribution in the series of instanton numbers 
for a given deformation parameter.
\end{abstract}


\newpage
\section{Introduction}

In 2009, a remarkable relation, so called ``AGT conjecture''\cite{Alday:2009aq}  was proposed between the Nekrasov partition function of  $ \mathcal{N}=2$ $SU(2)$ gauge theory in 4D 
\cite{hep-th/0206161,hep-th/0306238}
and the conformal block of the Liouville theory in 2D, 
relating  the $\Omega$-background parameters $\epsilon_1$ and $\epsilon_2$ and instanton expansion parameters with background charge of the Liouville theory and modular parameter of the conformal block.  In addition, the vacuum expectation value of the Coulomb branch parameter (VEV of the adjoint scalar) is related with the momentum of the primary field of the  intermediate  channel.   Soon after this relation was generalized to $SU(N)$ gauge group in \cite{arXiv:0907.2189} and \cite{Mironov:2009by, Bonelli:2009zp}.

After this conjecture,  various works have been performed using the Selberg integral and Jack polynomials  such as in  \cite{Itoyama:2010ki,arXiv:1003.5752,Itoyama:2010na,Alba:2010qc,arXiv:1012.3137,arXiv:1110.5255} and references therein. The Nekrasov partition function is obtained in the limit  $\epsilon_1+\epsilon_2=0$ corresponding to the $c=1$ Liouville theory 
(which is also called $\beta=1$ limit, which will be elaborated later). 
 Nevertheless, the conjecture seems to go beyond this limit $\beta \ne 1$ 
and the  Selberg integral provides a nice tool to this approach.  

Similarly  related but a little different approach we are going to investigate in this paper 
is to view the partition function 
in terms of  Penner type matrix model. This was initially proposed in \cite{Dijkgraaf:2009pc}  for four flavor case and generalized to less flavor cases in \cite{Eguchi:2009gf}
noting that the Liouville conformal block can be reproduced in terms of the $\beta$-deformation of hermitian  matrix model. In this matrix model,  the gauge parameters  and the $\Omega$-background parameters determines  the matrix couplings,  the deformed parameter  $\beta$ and the size of the matrix  $N$. 
These matrix models and related topics have widely been studied in \cite{arXiv:0911.4244}-\cite{arXiv:1106.1539}.

The free energy of the $\beta$-deformed matrix model is generally expanded in powers of the coupling $g$ corresponding to the genus expansion 
\begin{eqnarray}
\label{genus_expansion}
F \equiv 4g^2 \log Z= \sum_{n=0}^\infty F_n(\Lambda)\,g^n,
\label{eq:general_exp_free-energy}
\end{eqnarray}
where $\Lambda$ is a parameter involved in the matrix potential and interpreted as a dynamical scale of the corresponding gauge theory.\footnote{When the corresponding gauge theory has four flavors, $\Lambda$ should be identified with the exponential of a UV gauge coupling.}
When  $\beta=1$, the genus expansion terms  $F_{n}$ are vanishing for odd $n$. 
The planar free energy $F_{0}(\Lambda)$ was shown to be equivalent to the Seiberg-Witten prepotential \cite{Eguchi:2009gf, Itoyama:2010ki, Eguchi:2010rf}, and the genus one correction $F_{1}(\Lambda)$ was also shown to be consistent with the Nekrasov partition function \cite{arXiv:0912.2988}. 

When $\beta\neq 1$,  the genus expansion terms $F_{n}(\Lambda)$ in (\ref{genus_expansion})
is not vanishing also for odd $n$.  This is easily seen in the loop equation of  the $\beta$-deformed one  such as given in \cite{Chekhov-Eynard}. In this paper, we apply the usual loop equation technique to the
relatively simple model, namely, the  $\beta$-deformed matrix models for $N_F=2,3,4$ for $SU(2)$ gauge group case  and evaluate the half-genus correction $F_{1}(\Lambda)$, closely following the method employed in \cite{Eguchi:2010rf} but generalizing to $\beta\neq 1$ case and evaluate $F_{1}(\Lambda)$.   In fact, the loop equation was initially studied in \cite{arXiv:1104.2738} and only an integral expression is presented.  In this paper, we will calculate the explicit expression 
of the first half-genus correction to the free energy and compare the result with the corresponding Nekrasov partition function with general $\Omega$-background, including higher instanton contributions.  

The rest of this paper is organized as follows. In section \ref{sec:2}, we briefly review the matrix models for $N_F=2,3$ and $4$, and evaluate the resolvent of the matrix models using the loop equation
for $\beta$-deformed case.  
At the planar limit, the filling fraction is identified with the Coulomb branch parameter. 
However, it is noted that the relation is to be modified at the order of $\mathcal{O}(g)$. 
In section \ref{sec:3}, we concentrate on the $N_F=2$ case and evaluate the half-genus correction to the free energy. In section \ref{sec:4} and \ref{sec:5}, we generalize the previous argument to the $N_F=3$ and $4$ cases, respectively.  Section 6 is the summary and discussion.  
In appendix A, Penner type matrix model is constructed from the AGT conjecture. 
In appendix B, the derivation of the loop equation is reviewed for general $\beta$. 
In appendix C, the explicit expression for the Nekrasov partition function is shown for $N_F=2,3$ and $4$ for comparison with the matrix model results.


\section{Penner type models and half-genus corrections}
\label{sec:2}

In this section, we describe the Penner type matrix models proposed in \cite{Dijkgraaf:2009pc, Eguchi:2009gf}
to set up our approach.   Then, we solve the loop equation and find the expressions of the resolvent
for the planar and half-genus correction.  In addition, the Coulomb branch parameter is given in terms of filling fraction. 

\subsection{Penner type models for $SU(2)$ gauge theories}

Penner type matrix model was  proposed in \cite{Dijkgraaf:2009pc} for the $N_F=4$ case and later generalized in \cite{Eguchi:2009gf} to $N_F=2,3$ cases. 
The partition function $Z_{\rm matrix}$ of the matrix model is defined as 
\begin{eqnarray}
Z_{\rm matrix} = \left(\prod_{I=1}^N\int d\lambda_I\right) 
\Delta_N^{2\beta}\exp \left[\frac{\sqrt{\beta}}{g}\sum_{I}V(\lambda_I)\right],
\label{eq:general_partition}
\end{eqnarray}
where $g$ is a coupling constant of the matrix model, $\beta$ is a deformation parameter
and $\Delta_N =\prod_{I<J}(\lambda_I-\lambda_J)$ is the Vandermonde determinant.
When $\beta=1$, this reduces to the usual hermitian matrix model with a coupling $g$
and  $\lambda_I$ is the eigenvalues of the hermitian matrix. 

According to the AGT conjecture, the parameters $g$ and $\beta$  are 
related to the $\Omega$-background parameters $\epsilon_1, \epsilon_2$ of the gauge theory%
\footnote{We use the notation  $g$ as one half of the topological string coupling $g_s$.}
\begin{eqnarray}
 \epsilon_1 = 2g\sqrt{\beta},\qquad \epsilon_2 = -\frac{2g}{\sqrt{\beta}}.
\label{eq:relation1}
\end{eqnarray}
The matrix size $N$ is identified with the number of screening charges 
in Liouville theory and is related to the  the mass parameters of the gauge theory.
The details depend on the number of flavors  as explained below.

The free energy of the matrix model is defined by
\begin{eqnarray}
 F^{\rm matrix} \equiv 4g^2\log Z_{\rm matrix} = (-\epsilon_1\epsilon_2)\log Z_{\rm matrix} ,
\label{eq:free_energy} 
\end{eqnarray}
and is expanded in powers of $g$ as
\begin{eqnarray}
F^{\rm matrix} = F_{0}^{\rm matrix} + \frac{\epsilon_+}{2}F_{1}^{\rm matrix} + \mathcal{O}(g^2),
\end{eqnarray} 
 where $\epsilon_+=\epsilon_1+\epsilon_2$ is the order of  $g$ and 
the half-genus correction $F_{1}^{\rm matrix}$ is of our chief concern. In comparison to $F_1(\Lambda)$ in \eqref{eq:general_exp_free-energy}, we see $gF_1 = \frac{\epsilon_+}{2}F_1^{\rm matrix}$.

The explicit form of the potential $V(z)$ depends on the number of flavors $N_F$  in gauge theory. When $N_F=4$, the potential is given by 
\begin{eqnarray}
 V (z) &=& \left(m_0+\frac{\epsilon_+}{2}\right)\log z + m_1\log(z-1) + m_2\log(z-q),
\label{eq:pot_NF=4}
\end{eqnarray}
where $q$ in \eqref{eq:pot_NF=4} is identified with 
the exponential of a UV marginal coupling in the gauge theory.  
The mass parameters $m_0,m_1$ and $m_2$  (and with additional  $m_\infty$)
are associated to the Cartan sub-algebra of $SO(8)$ flavor symmetry 
and are related to the masses $\mu_I$ of the four {\em anti-fundamental} hypermultiplets by
\begin{eqnarray}
 \mu_1=m_1+m_\infty,\quad \mu_2=m_1-m_\infty,\quad \mu_3=m_2+m_0,\quad \mu_4=m_2-m_0.
\label{eq:mass_Penner}
\end{eqnarray}

We put potential term proportional to $\epsilon_+$ in \eqref{eq:pot_NF=4} 
so that the mass parameter relation in \eqref{eq:mass_Penner} is maintained
(see appendix \ref{app:AGT}). 
The matrix size $N$ is determined by the relation
\begin{eqnarray} 
\mu_1 + \mu_3 + 2\sqrt{\beta}g N = 0. 
\label{eq:relation2}
\end{eqnarray} 
which corresponds to the neutrality condition
in the presence of the background charge of the Liouville theory.
At the planar limit as explicitly shown in \cite{Dijkgraaf:2009pc}
the $\epsilon_+$ dependent correction can be neglected.

The  case with $N_F=3$ is obtained
if one takes the limit $\mu_4\to\infty$ while keeping $\Lambda_3 \equiv\mu_4q$ finite
so that  a single hypermultiplet is decoupled \cite{Gaiotto0908.0307, Eguchi:2009gf}. 
The resulting potential is given as  (neglecting a divergent constant term)
\begin{eqnarray}
V (z) &=& \left(\mu_3 + \frac{\epsilon_+}{2}\right)\log z + m_1\log(z-1) - \frac{\Lambda_3}{2z}.
\label{eq:pot_NF=3}
\end{eqnarray}
Here $\Lambda_3$ is a dimensionful parameter, and identified with the dynamical scale of the resulting gauge theory. 
The potential for $N_F=2$ is obtained  from $N_F=3$ case further by taking the  limit $\mu_2\to \infty$
 with $(\Lambda_2)^2\equiv \mu_2\Lambda_3$ fixed.%
\footnote{The sign of $\Lambda_2$ is different from that in \cite{Eguchi:2009gf}, but it is just the matter of convention.}
\begin{eqnarray}
 V(z) &=& \left(\mu_3+\frac{\epsilon_+}{2}\right)\log z + \frac{\Lambda_2}{2}\left(z + \frac{1}{z}\right).
\label{eq:pot_NF=2}
\end{eqnarray}
 Note that the relation \eqref{eq:relation2}  is not  changed by the limiting process 
since $\mu_1$ and $\mu_3$ are required to be finite in this limit. 

\subsection{loop equation and spectral curve}

We now discuss half-genus correction to the resolvent of the Penner type matrix models. 
The resolvent is defined by
\begin{eqnarray}
 W(z) &\equiv& \sqrt{\beta}g\left\langle\sum_{I}\frac{1}{z-\lambda_I}\right\rangle,
\label{eq:resolvent}
\end{eqnarray}
which can be expanded in powers of $g$:
\begin{eqnarray}
 W(z) = \sum_{n=0}^\infty \widetilde{W}_{n}(z)\,g^n
\label{eq:expansion_resolvent}
\end{eqnarray} 
The resolvent satisfies the $\beta$-deformed version of the loop equation \cite{Chekhov-Eynard, Chekhov, CEM}.
For our purpose of studying the half-genus correction,
we may neglect $\mathcal{O}(g^2)$ (see appendix \ref{app:loop_equation}) 
to get 
\begin{eqnarray}
W(z)^2 +\frac{\epsilon_+}{2}W'(z) + W(z)V'(z) - \frac{f(z)}{4} &=& 0,
\label{eq:general_loop}
\end{eqnarray}
where $f(z)$ is defined by
\begin{eqnarray}
 f(z) &\equiv& 4\sqrt{\beta}g\left\langle \sum_{I} \frac{V'(z)-V'(\lambda_I)}{z-\lambda_I}\right\rangle.
\label{eq:f}
\end{eqnarray}
It is noted that $\widetilde{W}_{n}(z)$ for odd $n$ does not vanishing when $\beta\neq 1$, 
resulting in the ``half-genus expansion'', rather than the usual genus-expansion for the case of $\beta=1$. 

To solve the loop equation \eqref{eq:general_loop}, we divide the potential into two parts \cite{arXiv:1104.2738}:
\begin{eqnarray}
 V(z) = V_0(z) + \frac{\epsilon_+}{2} V_1(z)
\label{eq:expansion_pot}
\end{eqnarray}
with  $V_0(z)$ and $V_1(z) = \log z,$  $\mathcal{O}(g^0)$. 
Putting the resolvent of the form 
\begin{eqnarray}
W(z) &=& W_{0}(z) + \frac{\epsilon_+}{2} W_{1}(z) + \mathcal{O}(g^2),
\label{eq:expansion_resolvent2}
\end{eqnarray}
one has 
\begin{eqnarray}
 (W_{0})^2 + V_0'W_{0} - \frac{f}{4} &=& 0,
\label{eq:leading_loop}
\\[2mm]
(2W_{0} + V_0') W_{1} + W_{0}' + W_{0}V_1' &=& 0,
\label{eq:sub_loop}
\end{eqnarray}
whose solution is given as 
\begin{eqnarray}
\label{eq:planar_resolvent} 
 W_{0}(z) &=& \frac{-V_0'(z) + \sqrt{V_0'(z)^2 + f(z)}}{2},
\\
\label{eq:half-genus_resolvent}
 W_{1}(z) &=& -\frac{W_{0}'(z) + W_{0}(z)V_1'(z)}{2W_{0}(z) + V_0'(z)}.
\end{eqnarray}

Now, the resolvent defines the spectral curve 
\begin{eqnarray}
x \;\equiv\; 2W(z) + V'(z) = x_0 + \frac{\epsilon_+}{2} x_1 + \mathcal{O}(g^2),
\label{eq:def_curve}
\end{eqnarray}
where $x_0$ is the  the planar contribution  
\begin{eqnarray}
 (x_0)^2 =  V_0'(z)^2 + f(z) 
\label{eq:x_0}
\end{eqnarray}
and  $x_1$ is the half-genus correction 
\begin{eqnarray}
x_1 = \frac{-x_0'(z)+V_0''(z) +V_0'(z)V_1'(z)}{x_0(z)}.
\label{eq:x_1}
\end{eqnarray}

The above Penner type potentials present two cuts in the  spectral curve
and it is proposed in  \cite{Dijkgraaf:2009pc}
that the Coulomb branch parameter $a$ of the gauge theory
is to be identified with  the ``filling fraction'' of the matrix model
\begin{eqnarray}
a = \frac{1}{2\pi i}\oint_{\mathcal{A}} xdz,
\label{eq:A-period}
\end{eqnarray}
where $\mathcal{A}$ denotes a cycle surrounding a branch cut associated to $x$. 
This proposal is explicitly checked in \cite{Eguchi:2010rf} for $\beta=1$ 
and  \eqref{eq:A-period} reproduces a correct UV behavior 
 $a\sim \sqrt{u}$ where $u$ denotes the VEV of the adjoint scalar in gauge theory.

On the other hand, the half-genus corrections of the matrix model
forces one to modify the identification into 
\begin{eqnarray}
 a = \frac{1}{2\pi i}\oint_{\mathcal{A}} xdz + \frac{\epsilon_+}{2}. 
\label{eq:Coulomb_general}
\end{eqnarray} 
This modification is necessary to keep the asymptotic behavior $a\sim \sqrt{u}$ in the UV limit of the gauge theory.   This is because the half-genus correction of the filling fraction 
does not vanish as $u\to \infty$ and the second term in \eqref{eq:Coulomb_general} 
cancels the non-vanishing half-genus correction. 

To elaborate on this modification, 
we redefine $a = a_0 + \frac{\epsilon_+}{2} a_1 + \mathcal{O}(g^2)$,
so that 
\begin{eqnarray}
a_0 \equiv \frac{1}{2\pi i}\oint_{\mathcal{A}} x_0 dz,\qquad a_1 \equiv \frac{1}{2\pi i}\oint_{\mathcal{A}} x_1 dz + 1.
\label{eq:a0a1_general}
\end{eqnarray}
Here $a_1$ can be put into more useful form if one uses  \eqref{eq:x_1}
\begin{eqnarray}
 a_1 = -\frac{1}{2\pi i}\oint_{\mathcal{A}}d(\log x_0) + \frac{1}{2\pi i}\oint_{\mathcal{A}}\frac{V_0''(z)+V_0'(z)V_1'(z)}{x_0(z)}dz +1.
\label{eq:pre-a1_general}
\end{eqnarray}
The first term gives just $-1$ due to the monodromy of the logarithmic function,
which encircles a square-root branch cut associated with $x_0$.
This enforces one to modify the identification of the Coulomb branch parameter \eqref{eq:Coulomb_general} so that this contribution is canceled out.  
Thus, one has the expression
\begin{eqnarray}
a_1 =  \frac{1}{2\pi i}\oint_{\mathcal{A}}\frac{V_0''(z)+V_0'(z)V_1'(z)}{x_0(z)}dz.
\label{eq:a1_general}
\end{eqnarray}

In the next sections, we evaluate the free energy of the matrix models by using $a_0$ and $a_1$
and compare with  the Nekrasov partition function for $N_F=2,3,4$ cases. 


\section{$N_F=2$ case}
\label{sec:3}

The potential \eqref{eq:pot_NF=2} (omitting the subscript of $\Lambda_2$ for simplicity)
is given as 
\begin{eqnarray}
V_0(z) = \mu_3 \log z + \frac{\Lambda}{2}\left(z + \frac{1}{z}\right),\qquad V_1(z) = \log z.
\label{eq:divide_potential}
\end{eqnarray} 
To find the spectral curve we evaluate  $f(z)$ \eqref{eq:f} in form 
 \cite{Eguchi:2009gf, Eguchi:2010rf}
\begin{eqnarray}
 f(z) &=& \frac{c_1}{z} + \frac{c_2}{z^2},
\end{eqnarray}
where $c_1$ and $c_2$ are given by
\begin{eqnarray}
 c_1 = 2\sqrt{\beta}gN\Lambda = -(\mu_1+\mu_3)\Lambda,
\qquad c_2 = 2\sqrt{\beta}g\sum_{I=1}^N \left\langle \frac{\Lambda}{\lambda_I}\right\rangle.
\label{eq:c_NF=2}
\end{eqnarray}
The explicit form of $c_1$ is obtained due to the relation \eqref{eq:relation2}. 
Since $c_1$ has no half-genus correction, the planar spectral curve has the same expression as in the $\beta=1$ case \cite{Eguchi:2010rf}:
\begin{eqnarray}
 (x_0)^2  \;=\; \left(\frac{\mu_3}{z}+\frac{\Lambda}{2}(1-\frac{1}{z^2})\right)^2 + \frac{c_1}{z} + \frac{c_2}{z^2} \;=\; \frac{\Lambda^2}{4}\frac{P_4(z)}{z^4},
\end{eqnarray}
where $P_4(z)$ is a polynomial of degree four of the form
\begin{eqnarray}
 P_4(z) &=& z^4 - \frac{4\mu_1}{\Lambda}z^3 + \frac{4}{\Lambda^2}\left(\mu_3^2 + c_2 - \frac{\Lambda^2}{2}\right)z^2  - \frac{4\mu_3}{\Lambda}z + 1.
\end{eqnarray}
This shows that the spectral curve is parameterized by a single complex paramter $c_2$, which will be interpreted as the moduli parameter of the gauge theory \cite{Eguchi:2009gf, Eguchi:2010rf}.
Hereafter, we set $\mu_1=\mu_3= m$ just for simplicity. Then $P_4(z)$ becomes
\begin{eqnarray}
 P_4(z) &=& z^4 - \frac{4m}{\Lambda}z^3 + \frac{4}{\Lambda^2}\left(m^2 + c_2 - \frac{\Lambda^2}{2}\right)z^2  - \frac{4m}{\Lambda}z + 1.
\end{eqnarray}

The Coulomb branch parameter $a$ is obtained from the 
filling fraction \eqref{eq:Coulomb_general}. 
The planar contribution $a_0$ was evaluated in \cite{Eguchi:2010rf} 
using the series expansion of hypergeometric function and has the form 
\begin{eqnarray}
a_{0} &=& \sqrt{A}\left(1-\frac{m^2}{4A^2}\Lambda^2 - \frac{(A^2 - 6m^2A + 15m^4)}{64A^4}\Lambda^4 - \frac{5(3m^2 A^2 - 14m^4A + 21 m^6)}{256A^6}\Lambda^6\right.
\nonumber \\[2mm]
&& \qquad \left. - \frac{15(A^4 - 28m^2A^3 + 294 m^4 A^2 - 924m^6A + 1001m^8)}{16384A^8}\Lambda^8 + \mathcal{O}(\Lambda^{10})
\right)
\label{eq:planar_filling}
\end{eqnarray}
where $A\equiv m^2 + c_2 - \frac{\Lambda^2}{2}$. 
The half-genus correction $a_1$ is easily obtained if one notices 
the relation%
\footnote{Similar structure has been noticed in \cite{Maruyoshi:2010iu} in the semi-classical approach of the system for each  $N_F$'s. }
$a_1 = -\left.\frac{\partial a_0}{\partial m}\right|_{A:\,{\rm fixed}}$
because $a_0$ and $a_1$ have the form 
\begin{eqnarray}
a_0 = \frac{1}{2\pi i}\oint_{\mathcal{A}} \frac{\Lambda}{2z^2}\sqrt{P_4(z)}dz
\\
a_1 = \frac{1}{2\pi i}\oint_{\mathcal{A}}\left(\frac{1}{z} + z\right)\frac{dz}{\sqrt{P_4(z)}}.
\label{eq:middle_a1}
\end{eqnarray}
This follows from  \eqref{eq:a1_general} and  \eqref{eq:divide_potential}. 
Therefore, we finally obtain
\begin{eqnarray}
a_1 &=& \frac{m}{2 A^{3/2}}\Lambda^2-\frac{3m \left(A-5 m^2\right)}{16 A^{7/2}}\Lambda^4+\frac{5 \left(3 A^2 m-28 A m^3+63 m^5\right)}{128 A^{11/2}} \Lambda^6
\nonumber\\[3mm]
&& \quad -\frac{105 \left(m \left(A^3-21 A^2 m^2+99 A m^4-143 m^6\right)\right) }{2048 A^{15/2}}\Lambda^8 + \cdots
\label{eq:sub_filling}
\end{eqnarray}
Note that $a_1$ vanishes in the limit of $A\to\infty$ so that $a \sim \sqrt{A}$ asymptotically. 
This is natural because  $A\equiv m^2 + c_2 - \frac{\Lambda^2}{2}$ is identified with the VEV of the adjoint scalar  \cite{Eguchi:2010rf} and the limit $A\to \infty$ corresponds to the UV limit in gauge theory.

The free energy \eqref{eq:free_energy} can be obtained using the relation  
\begin{eqnarray}
\Lambda\frac{\partial}{\partial \Lambda}F^{\rm matrix} \;=\; 2g\sqrt{\beta}\Lambda\sum_{I}\left\langle\lambda_I + \frac{1}{\lambda_I}\right\rangle \;=\; c_2 + 2g\sqrt{\beta}\Lambda\langle\sum_I\lambda_I\rangle.
\label{eq:general_free}
\end{eqnarray}
Noting the asymptotic behavior of the resolvent 
$W(z) \sim \sqrt{\beta}g N /z + \sqrt{\beta}g \langle \sum_I\lambda_I\rangle/z^2 + \mathcal{O}(1/z^3)$ 
one finds the terms of order $\mathcal{O}(z^{-2})$ in the loop equation
\eqref{eq:general_loop}  
\begin{eqnarray}
 2g\sqrt{\beta} \langle\sum_I\lambda_I\rangle &=& c_2 + (\mu_3^2-\mu_1^2) + \mathcal{O}(g^2).
\end{eqnarray}
Since we put  $\mu_1=\mu_3=m$,  
\eqref{eq:general_loop} reduces to
\begin{eqnarray}
 \Lambda\frac{\partial }{\partial \Lambda}F^{\rm matrix} \;=\; 2c_2 + \mathcal{O}(g^2)\;=\; 2(A -m^2) +\Lambda^2 + \mathcal{O}(g^2).
\label{eq:planar_free-2}
\end{eqnarray}

To integrate the right-hand side with respect to $\Lambda$, one needs to find the 
explicit form of $A$. Using the filling fraction 
$a= a_0 + \frac{\epsilon_+}2  a_1 $ given in 
\eqref{eq:planar_filling} and \eqref{eq:sub_filling}
one finds 
\begin{eqnarray}
 A &=& \left(a^2 + \frac{m^2}{2a^2}\Lambda^2 + \frac{a^4-6 a^2 m^2+5 m^4}{32 a^6}\Lambda^4 + \frac{5 a^4 m^2-14 a^2 m^4+9 m^6}{64 a^{10}}\Lambda^6\right.
\nonumber\\[2mm]
&& \qquad \left.+ \frac{5 a^8-252 a^6 m^2+1638 a^4 m^4-2860 a^2 m^6+1469 m^8}{8192 a^{14}}\Lambda^8 + \cdots\right)
\nonumber\\[4mm]
&&+ \frac{\epsilon_+}{2}\left( - \frac{m}{a^2}\Lambda^2 + \frac{3a^2m-5m^3}{8a^6}\Lambda^4 -\frac{5 a^4 m-28 a^2 m^3+27 m^5}{32 a^{10}}\Lambda^6\right.
\nonumber\\[2mm]
&&\qquad  \left.+\frac{63 a^6 m-819 a^4 m^3+2145 a^2 m^5-1469 m^7}{1024 a^{14}}\Lambda^8 + \cdots\right)
\nonumber\\[4mm]
&& + \mathcal{O}(g^2).
\label{eq:A0}
\end{eqnarray}
Putting this result into \eqref{eq:planar_free-2}, 
we obtain the free energy of the form
$  F^{\rm matrix} 
= F_0^{\rm matrix} + \frac{\epsilon_+}{2}F_1^{\rm matrix} + \mathcal{O}(g^2)$
where  $F_0^{\rm matrix} $ is the planar contribution
\begin{eqnarray}
 F^{\rm matrix}_0(a,m) &=& 2(a^2-m^2)\log \Lambda + \frac{a^2+m^2}{2a^2}\Lambda^2+ \frac{a^4-6a^2m^2 + 5m^4}{64a^6}\Lambda^4
\nonumber\\[2mm]
&& + \frac{5a^4m^2 - 14a^2m^4 +9m^6}{192a^{10}}\Lambda^6
\nonumber\\[2mm]
&&+ \frac{5 a^8-252 a^6 m^2+1638 a^4 m^4-2860 a^2 m^6+1469 m^8}{32768 a^{14}}\Lambda^8 + \cdots
\nonumber\\
\end{eqnarray}
and  $F^{\rm matrix}_1$ is the half-genus correction
\begin{eqnarray}
F^{\rm matrix}_1(a,m) &=&   - \frac{m}{a^2}\Lambda^2 + \frac{3a^2m-5m^3}{16a^6}\Lambda^4 -\frac{5 a^4 m-28 a^2 m^3+27 m^5}{96 a^{10}}\Lambda^6
\nonumber\\[4mm]
&&\quad +\frac{63 a^6 m-819 a^4 m^3+2145 a^2 m^5-1469 m^7}{4096 a^{14}}\Lambda^8 + \cdots.
\end{eqnarray}
Here $ F^{\rm matrix}_0 $ 
is the same as evaluated in \cite{Eguchi:2010rf} 
and agrees with the Seiberg-Witten prepotential of the corresponding gauge theory, while
$F^{\rm matrix}_1$ perfectly agrees with the half-genus correction to the Nekrasov partition function 
with  parameters $\mu_1 = \mu_3 = m$. (The corresponding Nekrasov partition function is given 
in \eqref{eq:half_inst_NF=2} for comparison.)

\section{$N_F=3$ case}
\label{sec:4}

For $N_F=3$ case, the corresponding potential of the matrix model is given by \eqref{eq:pot_NF=3}
with the definition  \eqref{eq:expansion_pot} 
\begin{eqnarray}
V_0(z) = \mu_3 \log z + m_1 \log (z-1) - \frac{\Lambda}{2z},\qquad V_1(z) = \log z.
\end{eqnarray}
$f(z)$ has the following expression 
\begin{eqnarray}
 f(z) &=& \frac{c_1}{z} + \frac{c_2}{z-1} + \frac{c_3}{z^2},
\end{eqnarray}
and $c_1,c_2$ and $c_3$ are given by
\begin{eqnarray}
 c_1 &=& -4\sqrt{\beta}g \sum_{I=1}^N\left\langle \frac{\mu_3+\epsilon_+/2}{\lambda_I} + \frac{\Lambda}{2\lambda_I^2}\right\rangle,\quad c_2 = -4\sqrt{\beta}g\sum_{I=1}^N\left\langle\frac{m_1}{\lambda_I-1}\right\rangle,
\nonumber\\
 c_3 &=& -2\sqrt{\beta}g\sum_{I=1}^N \left\langle \frac{\Lambda}{\lambda_I}\right\rangle.
\end{eqnarray}

The symmetry of the partition function $\langle \sum_I V'(\lambda_I)\rangle=0$ leads to the condition
\begin{equation}
\label{eq:constraint_c1_c2}
c_1 + c_2=0.
\end{equation}
Another constraint on $c_2$ and $c_3$ follows from the asymptotic 
behavior of  the loop equation \eqref{eq:general_loop}  
where the resolvent has the asymptotic behavior
$W(z) = \sqrt{\beta}g N/z + \mathcal{O}(1/z^2)$.
Noting that the derivative of the potential is $V'(z) = (\mu_3 + m_1 + \epsilon_+/2)/z + \mathcal{O}(1/z^2)$ and $f(z) = (c_2 + c_3)/z^2 + \mathcal{O}(1/z^3)$ with $c_1 + c_2 =0$
one has the left-hand side of the loop equation  at the order of $\mathcal{O}(1/z^2)$ 
\begin{eqnarray}
 c_2 + c_3 = 4\sqrt{\beta}g N\left(\sqrt{\beta}g N + \mu_3 + m_1\right)
= m_\infty^2 - (\mu_3 + m_1)^2,
\label{eq:constraints_c2c3}
\end{eqnarray}
where in the second equality we used equation \eqref{eq:relation2}.
The two constraints 
\eqref{eq:constraint_c1_c2} and \eqref{eq:constraints_c2c3} allow
one independent parameter, for which we choose $c_3$.

The planar spectral curve has the form 
\begin{eqnarray}
 (x_0)^2 &=& \frac{P_4(z)}{4z^4(z-1)^2},
\end{eqnarray}
where $P_4(z)$ is a polynomial of degree four which is given by
\begin{eqnarray}
 P_4(z) &=& 4m_\infty^2z^4 - 4(B-m_1\Lambda-m_1^2+m_\infty^2)z^3
\nonumber\\
&&\quad + (4B-4m_1\Lambda+ \Lambda^2 -4\mu_3\Lambda)z^2 +2\Lambda(2\mu_3-\Lambda)z + \Lambda^2,
\end{eqnarray}
with $B = c_3 -\mu_3\Lambda +\mu_3^2$. 
Hereafter, for simplicity, we set $m_1=m_\infty=0$ and $\mu_3 = m$ as in \cite{Eguchi:2010rf}.
Then $P_4(z)$ becomes a third order polynomial of $z$ as
\begin{eqnarray}
 P_4(z) &=& -4Bz^3 + (4B +\Lambda^2 - 4m\Lambda)z^2 + 2\Lambda(2m-\Lambda)z + \Lambda^2.
\label{eq:N3_P4}
\end{eqnarray}

The Coulomb branch parameter identified as \eqref{eq:Coulomb_general}
has the planar contribution 
\begin{eqnarray}
 a_0 &=& -\sqrt{B}\Biggl(1 + \frac{m}{4B}\Lambda - \frac{B+3m^2}{64B^{2}}\Lambda^2 + \frac{m}{256B^3}(5m^2+B)\Lambda^3
\nonumber\\
&&\qquad\qquad - \frac{1}{16384B^4}(3B^2 + 30m^2B + 175m^4)\Lambda^4
\nonumber\\
&&\qquad\qquad  + \frac{m}{65536B^5}(9B^2 + 70m^2B + 441m^4)\Lambda^5 + \mathcal{O}(\Lambda^6)\Biggr)
\end{eqnarray}
and the half-genus correction $a_1$ \eqref{eq:a1_general} has the form
\begin{eqnarray}
a_1 = \frac{\Lambda}{2\pi i}\oint_A \frac{1}{\sqrt{P_4(z)}}\left(\frac{1}{z}-1\right)dz.
\label{eq:middle-integral_NF=3}
\end{eqnarray}
As in $N_F=2$ case, one has 
\begin{eqnarray}
a_1 =-\left.\frac{\partial a_0}{\partial m}\right|_{B} \;+\; \left.\Lambda\frac{\partial a_0}{\partial B}\right|_m.
\label{eq:a1}
\end{eqnarray}
by noting 
\begin{eqnarray}
\left.\frac{\partial a_0}{\partial m}\right|_{B:\,{\rm fixed}}\!\!\! &=& -\frac{\Lambda}{2\pi i}\oint_A \frac{ dz}{z\sqrt{P_4(z)}},
\\[2mm]
\left.\frac{\partial a_0}{\partial B}\right|_{m:\, {\rm fixed}}\!\!\! &=& -\frac{1}{2\pi i}\oint_A \frac{dz}{\sqrt{P_4(z)}},
\end{eqnarray}
where we treated $m$ and $B$ in \eqref{eq:N3_P4} as independent variables. 
Hence, $a_1$ is given as follows:
\begin{eqnarray}
 a_1 &=& -\frac{\Lambda }{4 \sqrt{B}}+\frac{m \Lambda ^2}{32 B^{3/2}}-\frac{\left(B+3 m^2\right) \Lambda ^3}{256 B^{5/2}}+\frac{\left(9 B m+25 m^3\right) \Lambda ^4}{4096 B^{7/2}}
\nonumber\\[3mm]
&&-\frac{\left(9 B^2+90 B m^2+245 m^4\right)}{65536 B^{9/2}} \Lambda^5 +\frac{m \left(75 B^2+490 B m^2+1323 m^4\right)}{524288 B^{11/2}}\Lambda^6 +\cdots.
\nonumber\\
\end{eqnarray}
Note that $a_1$ vanishes in the limit $B\to\infty$, which guarantees the asymptotic behavior 
$a \sim a_0 $. 

Now, we evaluate the free energy of the matrix model 
\begin{eqnarray}
 \Lambda\frac{\partial}{\partial \Lambda}F^{\rm matrix} \;=\; -2\sqrt{\beta}g\sum_I\left\langle \frac{\Lambda}{\lambda_I}\right\rangle \;=\; c_3
\;=\; B +m\Lambda - m^2.
\label{eq:derF_NF=3}
\end{eqnarray}
In order to integrate this, we solve the inverse function of $a(B,m)$ as 
\begin{eqnarray}
 B &=& \left(a^2 - \frac{m\Lambda}{2} + \frac{a^2+m^2}{32a^2}\Lambda^2 + \frac{a^4 -6m^2a^2 + 5m^4}{8192a^6}\Lambda^4  +\frac{5 a^4 m^2-14 a^2 m^4+9 m^6}{262144 a^{10}}\Lambda^6+ \cdots\right)
\nonumber\\[2mm]
&&+\frac{\epsilon_+}{2}\Biggl(-\frac{\Lambda}{2} -\frac{m}{16a^2}\Lambda^2 +\frac{3 a^2 m-5 m^3}{2048 a^6}\Lambda^4 -\frac{5 a^4 m-28 a^2 m^3+27 m^5}{131072 a^{10}}\Lambda^6+\cdots\Biggr)
\nonumber\\[2mm]
&& + \mathcal{O}(g^2).
\end{eqnarray}
By integrating \eqref{eq:derF_NF=3} one obtain the free energy of the form 
\begin{eqnarray}
F_0^{\rm matrix} &=& (a^2-m^2)\log\Lambda+\frac{m\Lambda}{2} + \frac{a^2 + m^2}{64a^2} \Lambda^2
\nonumber\\[2mm]
&& + \frac{a^4 -6m^2a^2 + 5m^4}{32768\,a^6}\Lambda^4 +\frac{5 a^4 m^2-14 a^2 m^4+9 m^6}{1572864\, a^{10}}\Lambda^6 + \cdots
\\ 
F^{\rm matrix}_1 &=&  - \frac{\Lambda}{2} - \frac{m}{32a^2}\Lambda^2 + \frac{3a^2m-5m^3}{8192\,a^6}\Lambda^4  -\frac{5 a^4 m-28 a^2 m^3+27 m^5}{786432\, a^{10}}\Lambda^6 
+ \cdots.
\nonumber\\
\end{eqnarray}
Here $F_0^{\rm matrix} $ was evaluated in \cite{Eguchi:2010rf} and is equal to the Seiberg-Witten prepotential of $N_F=3$ gauge theory. On the other hand, the half-genus correction $F_1^{\rm matrix}$ is newly evaluated here
and perfectly coincides with the half-genus correction to the Nekrasov partition function of $SU(2),\,N_F=3$ gauge theory  for our parameters $m_1 = m_\infty = 0,\,\mu_3 = m$ (See \eqref{eq:half_inst_NF=3} in appendix \ref{app:Nekrasov}).


\section{$N_F=4$ case}
\label{sec:5}

For $N_F=4$ gauge theory, the matrix model is 
given by \eqref{eq:pot_NF=4} with 
\begin{eqnarray}
 V_0(z) = m_0\log z + m_1 \log(z-1) + m_2\log(z-q),\qquad V_1(z) = \log z,
\end{eqnarray}
so that $V(z) = V_0(z) + (\epsilon_+/2)V_1(z)$. 
The function $f(z)$ defined in \eqref{eq:f} is now written as
$  f(z) = \sum_{i=0}^2{\frac{c_i}{z-q_i}}$ 
where $c_i$'s are given by
\begin{eqnarray}
 c_0 &=& -4\sqrt{\beta}g \sum_I\left\langle \frac{m_0 + \epsilon_+/2}{\lambda_I}\right\rangle,\quad c_1 \;=\; -4\sqrt{\beta}g \sum_{I}\left\langle \frac{m_1}{\lambda_I-1}\right\rangle,
\nonumber\\
c_2 &=& -4\sqrt{\beta}g \sum_I\left\langle \frac{m_2}{\lambda_I-q} \right\rangle.
\label{eq:c2_NF=4}
\end{eqnarray}

From the symmetry of the system $\sum_{I}\langle V'(\lambda_I)\rangle =0$, it follows that
\begin{eqnarray}
\sum_{i=0}^2c_i = 0.
\label{eq:const1_NF=4}
\end{eqnarray}
From the residues at infinity, we find an another constraint
\begin{eqnarray}
 c_1 + qc_2 = 4\sqrt{\beta}g N\left(\sqrt{\beta}gN +\sum_{i=0}^2m_i\right)
= m_\infty^2 - \left(\sum_{i=0}^2m_i\right)^2,
\label{eq:const2_NF=4}
\end{eqnarray}
where in the second equality \eqref{eq:relation2} is used. Thus, essentially there is a single free parameter, for which we take $c_0$.

The planar spectral curve is exactly what was obtained in \cite{Eguchi:2010rf}:
\begin{eqnarray}
(x_0)^2 = \frac{P_4(z)}{z^2(z-1)^2(z-q)^2},
\end{eqnarray}
where $P_4(z)$ is a degree four polynomial of $z$.\footnote{For the explicit expression of $P_4(z)$ for general mass parameters, see equation (3.39) in \cite{Eguchi:2010rf}.} 
Hereafter, setting $m_0 = m_\infty = 0$ and $m_1 = m_2 = m$  for simplicity
(which provides the equal mass for hypermultiplets $\mu_i=m$ for $i=1,\cdots,4$), 
one has $P_4(z)$ of the third order polynomial 
\begin{eqnarray}
 P_4(z) &=& Cz^3 + \left\{ (1-q)^2 m^2-C(1+q)\right\}z^2 + Cqz,
\end{eqnarray}
where we defined $C\equiv qc_0$. 

The Coulomb branch parameter is given by \eqref{eq:a0a1_general}.
The planar contribution $a_0$ was already calculated in \cite{Eguchi:2010rf}, which is written
in our notation as
\begin{eqnarray}
a_0 &=& i\sqrt{C}\left(h_0(q) - h_1(q)\frac{m^2}{C} - \frac{h_2(q)}{3}\frac{m^4}{C^2} - \frac{h_3(q)}{5}\frac{m^6}{C^3} -\frac{h_4(q)}{7} \frac{m^8}{C^4}+\mathcal{O}\left(\frac{m^{10}}{C^5}\right)\right),\label{eq:a0_NF=4}
\nonumber\\
\end{eqnarray}
where $h_i(q)$ is defined in terms of the expansion coefficients of a hypergeometric function \cite{Eguchi:2010rf}
\begin{eqnarray}
h_0(q) &=& 1+\frac{1}{4}q + \frac{9}{64}q^2 + \frac{25}{256}q^3 + \frac{1225}{16384}q^4 + \mathcal{O}(q^5),
\\[2mm]
h_1(q) &=& \frac{1}{2}+\frac{1}{8}q + \frac{1}{128}q^2 +\frac{1}{512}q^3 + \frac{25}{32768}q^4 + \mathcal{O}(q^5),
\\[2mm]
h_2(q) &=& \frac{3}{8} + \frac{27}{32}q + \frac{27}{512}q^2 + \frac{3}{2048}q^3 + \frac{27}{131072}q^4 + \mathcal{O}(q^5),
\\[2mm]
h_3(q) &=& \frac{5}{16}+\frac{125}{64}q + \frac{1125}{1024}q^2 + \frac{125}{4096}q^3 + \frac{125}{262144}q^4 + \mathcal{O}(q^5).
\end{eqnarray}

The half-genus correction $a_1$ is given by \eqref{eq:a1_general}
\begin{eqnarray}
a_1 &=& - \frac{m(1+q)}{2\pi i}\oint_A \frac{dz}{\sqrt{P_4(z)}} -  \frac{m(1-q)^2}{2\pi i}\oint_A \frac{dz}{\sqrt{P_4(z)}}\frac{z}{(z-1)(z-q)}.
\label{eq:a1_middle_NF=4}
\end{eqnarray}
This integral is  expressed in terms of $a_0$:
\begin{eqnarray}
 \left.\frac{\partial a_0}{\partial C}\right|_{m:\,{\rm fixed}} &=& \frac{1}{4\pi i}\oint_A \frac{dz}{\sqrt{P_4(z)}},
\\[2mm]
\left.\frac{\partial a_0}{\partial m}\right|_{C:\,{\rm fixed}} &=& \frac{m(1-q)^2}{2\pi i}\oint_A\frac{dz}{\sqrt{P_4(z)}}\frac{z}{(z-1)(z-q)}.
\end{eqnarray}
Hence, we finally obtain 
\begin{eqnarray}
 a_1 &=& 
 -\left.2m(1+q)\frac{\partial a_0}{\partial C}\right|_{m} -\,\left.\frac{\partial a_0}{\partial m}\right|_{C}
\nonumber\\ 
&=& i\left[g_1(q)\frac{m}{\sqrt{C}} + g_3(q) \frac{m^3}{C^{3/2}} + g_5(q)\frac{m^5}{C^{5/2}} + g_7(q)\frac{m^7}{C^{7/2}}+\cdots\right].
\label{eq:a1_NF=4}
\end{eqnarray}
where $g_i(q)$ are functions of $q$ defined in terms of $h_i(q)$. The first few components are given by
\begin{eqnarray}
 g_1(q) &=& (-h_0(q)+2h_1(q)-h_0(q)q),
\\[2mm]
 g_3(q) &=& \frac{4h_2(q)-3(1+q)h_1(q)}{3},
\\
 g_5(q) &=& \frac{6h_3(q) - 5(1+q)h_2(q)}{5},
\\
 g_7(q) &=& \frac{8h_4(q)-7(1+q)h_3(q)}{7}.
\end{eqnarray}

To find the free energy $F^{\rm matrix}$ of the matrix model  
one can use the equation
\begin{eqnarray}
\frac{\partial}{\partial q} F^{\rm matrix} = 4g\sqrt{\beta} m_2\left\langle {\rm Tr}\frac{1}{q-M}\right\rangle = \left.4m_2W(z)\right|_{z=q}.
\end{eqnarray}
From the loop equation \eqref{eq:general_loop}, we find that $W(z)$ in the vicinity of $z = q$ 
\begin{eqnarray}
 W(z) = \frac{c_2}{4m_2} + \mathcal{O}(z-q),
\end{eqnarray}
which implies that
\begin{eqnarray}
\frac{\partial}{\partial q} F^{\rm matrix} \;=\; c_2 \;=\; \frac{1}{(1-q)}\left(4m^2 -\frac{C}{q}\right).\label{eq:free-energy}
\end{eqnarray}
Here $C$ is obtained from $a(C)$ as
\begin{eqnarray}
 C(a) &=& -a^2\left(\frac{1}{h_0(q)^2} - \frac{2h_1(q)}{h_0(q)}\frac{m^2}{a^2} + \frac{2h_0(q)h_2(q) - 3h_1(q)^2}{3}\frac{m^4}{a^4}\right.
\nonumber \\[2mm]
&& \qquad\qquad \left. - \frac{10h_0(q)h_1(q)^3 -10h_0(q)^2h_1(q)h_2(q) + 2h_0(q)^3h_3(q)}{5}\frac{m^6}{a^6} + \cdots\right)
\nonumber\\[2mm]
&&+\frac{\epsilon_+m}{2}\left(\frac{2(1+q)h_0(q)-4h_1(q)}{h_0(q)} + \frac{8h_0(q)h_2(q)-12h_1(q)^2}{3}\frac{m^2}{a^2} \right.
\nonumber\\[3mm]
&&\qquad\qquad\left.- \frac{12h_0(q)\{5h_1(q)^3 - 5h_0(q)h_1(q)h_2(q) + h_0(q)^2h_3(q)\}}{5}\frac{m^4}{a^4} + \cdots
\right)
\nonumber\\
&& +\; \mathcal{O}(g^2).
\label{eq:expand_C}
\end{eqnarray}
This is used to integrate \eqref{eq:free-energy}
to get the free energy $F^{\rm matrix}$ as
\begin{eqnarray}
F_0^{\rm matrix} &=& (a^2-m^2)\log q + \frac{a^4 +6a^2 m^2 + m^4}{2a^2}q
\nonumber\\[2mm]
&& + \frac{\left(13 a^8+100 a^6 m^2+22 a^4 m^4-12 a^2 m^6+5 m^8\right)}{64 a^6}q^2
\nonumber\\[2mm]
&& + \frac{23 a^{12}+204 a^{10} m^2+51 a^8 m^4-48 a^6 m^6+45 a^4 m^8-28 a^2 m^{10}+9 m^{12}}{192 a^{10}}q^3
\nonumber\\[2mm]
&& + \frac{1}{32768 a^{14}}\left(2701 a^{16}+26440 a^{14} m^2+7164 a^{12} m^4-9000 a^{10} m^6\right.
\nonumber\\[2mm]
&&\qquad \left.+12190 a^8 m^8-13384 a^6 m^{10}+10908 a^4 m^{12}-5720 a^2 m^{14}+1469 m^{16}\right)q^4
\nonumber\\[2mm]
&& + \mathcal{O}(q^5),
\\[4mm]
F_1^{\rm matrix} &=& -\frac{2m(a^2+m^2)}{a^2}q - \frac{9 a^6 m+11 a^4 m^3-9 a^2 m^5+5 m^7 }{8 a^6}q^2
\nonumber\\[2mm]
&&-\frac{38 a^{10} m+51 a^8 m^3-72 a^6 m^5+90 a^4 m^7-70 a^2 m^9+27 m^{11}}{48 a^{10}}q^3
\nonumber\\
&&-\frac{1}{2048 a^{14}} \left(1257 a^{14}m+1791 a^{12} m^3-3375 a^{10} m^5+6095 a^8 m^7-8365 a^6 m^9\right.
\nonumber\\
&&\qquad\qquad\left. +8181 a^4 m^{11}-5005 a^2 m^{13}+1469 m^{15}\right)q^4\;+\;\mathcal{O}(q^5).
\end{eqnarray}
Here $F_0^{\rm matrix}$ was  calculated in \cite{Eguchi:2010rf} 
and shown to be equivalent to the Seiberg-Witten prepotential of $N_F=4$ superconformal gauge theory.
The newly found  half-genus correction $F_1^{\rm matrix}$ coincides with the half-genus correction to the corresponding Nekrasov partition function \eqref{eq:half_inst_NF=4} 
for $\mu_i=m$. 

\section{Summary and Discussions}

In this paper, we investigate the $\beta$-ensemble of the matrix model
known as Penner type matrix model.
By solving the loop equation generalized to  $\beta\neq 1$ 
we explicitly evaluate $F_1^{\rm matrix}$.
The result perfectly agrees with the Nekrasov partition function
with general $\Omega$-background parameters $\epsilon_1,\epsilon_2$,
corresponding to the $\mathcal{N}=2,SU(2)$ gauge theories with $N_F=2,3$ and $4$ flavors.

It is noted that the relation of the filling fraction with the Coulomb branch parameter 
is to be modified as in \eqref{eq:Coulomb_general} at the order of half-genus expansion,
which reduces to the original one proposed in \cite{Dijkgraaf:2009pc} in the planar limit. 
The modification is  understood as a condition of the UV limit of the gauge theory so that 
the the correction to the Coulomb branch parameter need to vanish at each order of (half-) genus expansion as the corresponding scale parameter becomes infinite.

The loop equation provides a systematic way of finding the (half-) genus expansion of the theory.  The merit of the expansion is that even at this first non-trivial order,  the higher instanton contribution can be obtained by the simple algebraic algorithm.  
To go to high genus expansion, one needs to solve the lower genus result of multi-point resolvent
and the result will be reported in the near future. 
In addition, the extension of the analysis to the multi-matrix model is highly desired 
so that the Selberg integral approach of the $SU(N)$ gauge theory  \cite{arXiv:1110.5255}  is to be compared.

\subsection*{Acknowledgments} 
The authors thank Yutaka Matsuo and Hiroaki Nakajima
for useful discussion. 
This work is partially  supported
by the National Research Foundation of Korea (NRF)
grant funded by the Korea government (MEST) 2005-0049409.

\appendix

\section{AGT conjecture and Matrix model}
\label{app:AGT}

We briefly review the AGT conjecture \cite{Alday:2009aq} and its relation with the Penner type matrix models  which is viewed as dual to $d=4, \mathcal{N}=2$ gauge theories \cite{Dijkgraaf:2009pc}
for the setup of our convention in the manuscript.  

\subsection{AGT relation}

 The authors of \cite{Alday:2009aq} pointed out that Nekrasov partition functions of $d=4,\mathcal{N}=2, SU(2)$ gauge theories is identified with conformal block of of the  four-point correlation function of the Liouville theory,
\begin{eqnarray} 
&&
\langle V_{\widetilde{m}_\infty + \frac{Q}{2}}(\infty)V_{\widetilde{m}_1}(1)V_{\widetilde{m}_2}(q)V_{\widetilde{m}_0+\frac{Q}{2}}(0)\rangle_{\rm Liouville} 
\nonumber\\
&& ~~~~
= c(\mu_I,\epsilon_i)~
\left|q^{Q^2/4 -\Delta_{\widetilde{m}_2}-\Delta_{\widetilde{m}_0+\frac{Q}{2}}}\right|^2
~ \int a^2 da \left|Z_{\rm Nekrasov}^{SU(2)}(a,\mu_I,\epsilon_i)\right|^2
\label{eq:AGT}
\end{eqnarray}
where $V_{\alpha}(z) = e^{2\alpha\phi(z)}$ is a vertex operator of dimension $\Delta_\alpha = \alpha(Q-\alpha)$.  
$Q=b+1/b$ is the Liouville background charge and $b$ is the parameter in the Liouville potential $e^{2b\phi}$.
 $Z_{\rm Nekrasov}^{SU(2)}(a,\mu_I,\epsilon_i)$ in the right-hand side  is the Nekrasov partition function of $\mathcal{N}=2,N_F=4,SU(2)$ gauge theory
and depends on the Coulomb branch parameter $a$, masses of four hyper multiplets $\mu_I$, and the $\Omega$-background parameters $\epsilon_1$ and $\epsilon_2$. 
The prefactor $c(\mu_I,\epsilon_i)$ depends only on $\mu_I$ and $\epsilon_i$. 

 $q$ in \eqref{eq:AGT} is  the exponential of the UV gauge coupling and the relation \eqref{eq:AGT} is complete if  the Liouville parameters are given in terms of the gauge field parameters. 
 If one introduces a mass scale $\hbar$  (which was set to be 1 in \cite{Alday:2009aq})
so that $ \hbar \widetilde{m}_a \equiv m_a $, one has 
\begin{equation}
{\mu_1} = {m}_1 + {m}_\infty,
~~ {\mu_2} = {m}_1 - {m}_\infty,
~~ {\mu_3}  = {m}_2 + {m}_0,
~~{\mu_4} = {m}_2- {m}_0 \,.
\label{eq:mass_AGT}
\end{equation}

$\Omega$-background parameters $\epsilon_i$ are identified with the Liouville parameter $b$ as
\begin{equation}
 \epsilon_1 = \hbar b,\qquad \epsilon_2 = {\hbar}/{b},
\label{eq:epsilon_hbar}
\end{equation}
so that the background charge is written as  
$ Q = ({\epsilon_1 + \epsilon_2})/{\hbar} = {\epsilon_+}/{\hbar}$. 

\subsection{Dijkgraaf-Vafa's proposal of matrix model}

Inspired by the AGT relation, a matrix model description was proposed in \cite{Dijkgraaf:2009pc}. 
Note that the Liouville correlation can be evaluated perturbatively 
\begin{eqnarray}
 \left\langle \left(\prod_{I=1}^N\int d\lambda_Id\overline{\lambda}_I\; e^{2b\phi(\lambda_I)}\right)V_{\widetilde{m}_\infty+\frac{Q}{2}}(\infty)V_{\widetilde{m}_1}(1)V_{\widetilde{m}_2}(q)V_{\widetilde{m}_0+\frac{Q}{2}}(0)\right\rangle 
\label{eq:screening}
\end{eqnarray}
where the expectation values is evaluated in terms of the free field description
$\langle \phi(z)\phi(w) \rangle =  -1/2\log (z-w)^2$ 
or  $\langle e^{2\alpha_1\phi(z)}e^{2\alpha_2\phi(w)}\rangle  = |z-w|^{-4\alpha_1\alpha_2}$. 
The number $N$ of integrals comes from the perturbation of the Liouville potential
and is viewed as screening integrals.  The non-vanishing contribution is obtained if the number of screening charge satisfies the  neutrality condition
\begin{equation}
\sum_{i=0}^2 \widetilde{m}_i +\widetilde{m}_\infty + bN = 0.
\label{eq:screeningcondition}
\end{equation}
The correlation \eqref{eq:screening} is given as (up to the $q$-independent prefactor)
\[
 \left| q^{\frac{2(m_0+\epsilon_+/2)m_2}{\hbar^2}}(1-q)^{\frac{2m_1m_2}{\hbar^2}} ~I_4
\right|^2
\]  where 
\begin{eqnarray}
I_4 &=&   \int  \left[ \prod_{I=1}^Nd\lambda_I \right]
 ~ \prod_{I<J}\left(\lambda_I-\lambda_J\right)^{-2b^2} ~\exp\left(-\frac{2b}{ \hbar} \sum_I V(\lambda_I)\right) 
\label{eq:pre_matrix}
\\
V(z) &=& \left(m_0+ \frac{\epsilon_+}{2}\right)\log z + m_1\log (z-1) + m_2\log(z-q) .
\end{eqnarray} 

Note that  $I_4$ is not well defined unless  the integration ranges 
and parameters  $b$ and $\hbar$  are to be appropriately arranged to make  $I_4$ convergent.  
To fix this problem, one may consider the integrals with
$b^2= -\beta^2$ and $\hbar b = 2g \sqrt{\beta}$ (or $b = i\sqrt{\beta}, ~\hbar = -2ig$)
so that integration is well defined even when any two of the integration variables coincide.   
This defines the partition function of matrix model 
\begin{eqnarray}
Z_{\rm matrix}  \equiv \int  \left[ \prod_{I=1}^Nd\lambda_I \right]
 ~ \prod_{I<J}\left(\lambda_I-\lambda_J\right)^{2 \beta^2} ~\exp\left(\frac{\sqrt{\beta}}{ g} \sum_I V(\lambda_I)\right) 
\label{eq:penner}
\end{eqnarray} 
where $\lambda_I$ is the eigenvalue of the hermitian matrix and 
the size of the matrix $N$ is given from the neutrality condition 
\eqref{eq:screeningcondition} or  \eqref{eq:relation2}.
However, the matrix model is not the usual one unless $\beta=1$ and 
is called $\beta$-deformed matrix model, or Penner type matrix model.

As seen in section \ref{sec:2}, $Z_{\rm matrix} $  depends on a single parameter 
which is not in the potential $V(z)$ and the parameter can be chosen as the filling fraction. 
The origin of the ambiguity comes from $I_4$ in \eqref{eq:pre_matrix}
where one needs to arrange the integration range appropriately,
whose origin also traces back to the perturbation expansion of the Liouville correlation 
\eqref{eq:screening}.   
This ambiguity does not appear if one evaluate the Liouville correlation 
 using the conformal block  as put in \eqref{eq:AGT}. 
To resolve this discrepancy, one may view the perturbation 
as the one with the fixed filling fraction so that the integration ranges are chosen 
so that $N_1$ number of integration ranges from $0$ to $q$ and 
$N-N_1$ number of integration ranges from $1$ to $\infty$
resulting the filling fraction is related with the $N_1/N$.

According to AGT, VEV  of $SU(2)$ gauge group 
is identified with the Liouville momentum  $a$ of the conformal block 
in \eqref{eq:AGT}.  On the other hand, the Seiberg-Witten curve is ``quantized'' and 
the VEV is identified with the filling fraction of the S-W curve.
In this sense, it is very natural \cite{Dijkgraaf:2009pc}
that the Coulomb branch parameter of the gauge theory
is identified with the filling fraction of the matrix model. 
With this interpretation in mind, one may identify 
$ Z_{\rm matrix} $ with $Z_{\rm Nekrasov}$ as follows: 
\begin{eqnarray}
 Z_{\rm matrix} &=& q^{\frac{(m_0 - m_2)^2-2m_2\epsilon_+}{\hbar^2}}(1-q)^{-\frac{2m_1m_2}{\hbar^2}}Z_{\rm Nekrasov}^{\rm SU(2)}
\label{eq:matrix-SU2}.
\end{eqnarray}

On the other hand, it was pointed out in \cite{Alday:2009aq} that the Nekrasov partition functions of $SU(2)$ and $U(2)$ gauge theories are related by
\begin{eqnarray}
 Z_{\rm Nekrasov}^{SU(2)}(a,\mu_I,\epsilon_i) &=& (1-q)^{\frac{(\mu_1 + \mu_2)(\mu_3 + \mu_4)}{2\hbar^2}}Z_{\rm Nekrasov}^{U(2)}(\vec{a},\mu_I,\epsilon_i),
\label{eq:SU2-U2}
\end{eqnarray}
Note that the $U(2)$ gauge theory has two independent Coulomb branch parameters $\vec{a}=(a_1,a_2)$, but we set $a_1=-a_2 = a$ in the right-hand side of \eqref{eq:SU2-U2}. Note also that our $\mu_I$ are masses of {\em anti-fundamental} hyper multiplets, as states in appendix \ref{app:Nekrasov}. 
By combining \eqref{eq:matrix-SU2} and \eqref{eq:SU2-U2}, we obtain
\begin{eqnarray}
 Z_{\rm matrix} &=& q^{\frac{(m_0-m_2)^2-2m_2\epsilon_+}{\hbar^2}}Z_{\rm Nekrasov}^{U(2)},
\end{eqnarray}
up to a prefactor which is independent of $q$ and $a$.

\section{$\beta$-deformed version of loop equations}
\label{app:loop_equation}

We present the derivation of the loop equation for $\beta$-deformed Penner type models, following \cite{Chekhov} (See also \cite{Chekhov-Eynard, CEM}).
We start from the partition function of the form
\begin{eqnarray}
 Z = \int  \left[ \prod_{I=1}^N d\lambda_I\right]  \Delta_N^{2\beta} e^{\sqrt{\beta}/g\sum_{I=1}^N V(\lambda_I)}
\end{eqnarray}
and consider the change of the integration variable as $\lambda_I \to \lambda_I + \frac{\epsilon}{\lambda_I-z}$. This changes the expression of the integrand as well as the measure. Collecting terms proportional to $\epsilon$, we obtain
\begin{eqnarray}
0 &=& -\sum_{I,J=1}^N \left\langle \frac{\beta}{(\lambda_I-z)(\lambda_J-z)}\right\rangle - \sum_{I=1}^N \left\langle \frac{1-\beta}{(\lambda_I-z)^2}\right\rangle
\nonumber \\[2mm]
&&\quad  - \frac{\sqrt{\beta}}{g}V'(z)\sum_{I=1}^N\left\langle \frac{1}{z-\lambda_I}\right\rangle 
 + \frac{\sqrt{\beta}}{g}\sum_{I=1}^N\left\langle\frac{V'(z)-V'(\lambda_I)}{z-\lambda_I}\right\rangle.
\label{eq:pre_loop}
\end{eqnarray}
Here, the first and second terms come from the variations of the measure and $\Delta_N^{2\beta}$, while the third and fourth terms are from the variation of the potential $V(\lambda_I)$. By defining 
multi-point  (connected) resolvent as
\begin{eqnarray}
 W(z_1,\cdots,z_s) = \beta\left(\frac{g}{\sqrt{\beta}}\right)^{2-s}\left\langle \sum_{I_1}\frac{1}{z_1-\lambda_{I_1}}\cdots \sum_{I_s}\frac{1}{z_s-\lambda_{I_s}} \right\rangle_{\!\!\rm conn}
\end{eqnarray}
the equation \eqref{eq:pre_loop} gives 
 the loop equation for the $\beta$-deformed matrix model
\begin{eqnarray}
0 &=&  g^2  W(z,z)+W(z)^2 +g\left(\sqrt{\beta}-\frac{1}{\sqrt{\beta}}\right)W'(z) + V'(z)W(z) - \frac{f(z)}{4}
\label{eq:loop-equation0}
\\
 f(z) &=& 4g\sqrt{\beta}\sum_{I=1}^N\left\langle \frac{V'(z)-V'(\lambda_I)}{z-\lambda_I}\right\rangle.
\end{eqnarray}
Note that the first term in \eqref{eq:loop-equation0} is $\mathcal{O}(g^2)$, while the third term is  $\mathcal{O}(g)$ which contributes to the half-genus correction.  
Omitting higher order term, one has the loop equation up to  half-genus correction 
with $\epsilon_+=2g(\sqrt{\beta}-1/\sqrt{\beta})$
\begin{eqnarray}
 W(z)^2  +\frac{\epsilon_+}{2}W'(z) + V'(z)W(z) - \frac{f(z)}{4} = 0. 
\end{eqnarray}

\section{Nekrasov Partition Function}
\label{app:Nekrasov}

We present the Nekrasov partition function \cite{hep-th/0206161, hep-th/0306238} of $U(2)$ gauge theories for the comparison with the matrix result.  The Nekrasov partition function 
is factorized into the following three contributions:
\begin{eqnarray}
 Z = Z_{\rm class}Z_{\rm 1-loop}Z_{\rm inst},
\label{eq:factor_Nek}
\end{eqnarray}
where $Z_{\rm class}$ and $Z_{\rm 1-loop}$ are the classical and 1-loop contributions, respectively.
In the following, the instanton part is elaborated for $N_F =4,3,2$ cases. 

\subsection{$N_F=4$ theory}

\begin{eqnarray}
 Z_{\rm inst} = \sum_{\vec{Y}}q^{|\vec{Y}|}Z_{\rm vec}(\vec{a},\vec{Y})Z_{\rm afund}(\vec{a},\vec{Y},\mu_1)Z_{\rm afund}(\vec{a},\vec{Y},\mu_2)
 Z_{\rm afund}(\vec{a},\vec{Y},\mu_3)Z_{\rm afund}(\vec{a},\vec{Y},\mu_4),
\nonumber\\
\label{eq:inst_NF=4}
\end{eqnarray}
where the sum runs over pairs of Young diagrams $\vec{Y}=(Y_1,Y_2)$ and $\vec{a} = (a_1,a_2)$ denotes the Coulomb branch parameter while $\mu_i$ are mass parameters of four fundamentals. 
The vector multiplet contribution $Z_{\rm vec}(\vec{a},\vec{Y})$ is given by
\begin{eqnarray}
 Z_{\rm vec}(\vec{a},\vec{Y}) &=& \prod_{i,j=1}^2 \prod_{s\in Y_i}E(a_i-a_j,Y_i,Y_j,s)\prod_{t\in Y_j}(\epsilon_+ - E(a_j-a_i,Y_j,Y_i,t)),
\end{eqnarray}
where $s$ and $t$ run over all the boxes in $Y_i$ and $Y_j$, respectively.
The constituent $E(a,Y_i,Y_j,s)$ is defined by
\begin{eqnarray}
 E(a,Y_i,Y_j,s) = a - \epsilon_1 L_{Y_j}(s) + \epsilon_2(A_{Y_i}(s)+1).
\end{eqnarray}
The arm-length $A_{Y}(s)$ and leg-length $L_{Y}(s)$ are defined by
\begin{eqnarray}
 A_{Y}(s) = \lambda_k -l,\qquad L_{Y}(s) = \lambda_l'-k,
\end{eqnarray} 
where $(k,l)$ is the coordinate of the box $s$ in $Y$. The two integers $\lambda_k$ and $\lambda_k'$ denote the heights of $k$-th column of $Y$ and $Y^T$, respectively.

The anti-fundamental hyper multiplet contribution  is replaced by
\begin{eqnarray}
 Z_{\rm afund}(\vec{a},\vec{Y},\mu) &=& \prod_{i=1}^2\prod_{s\in Y_i}(\phi(a_i,s)+\mu),
\end{eqnarray}
where $\phi(a,s) \equiv a + \epsilon_1(k-1) + \epsilon_2(l-1)$ when $s$ is on the position $(k,l)$ in $Y$.
Note that in \cite{Eguchi:2010rf}  the last two factors in \eqref{eq:inst_NF=4} were given as \begin{eqnarray}
 Z_{\rm fund}(\vec{a},\vec{Y},-\mu_3)Z_{\rm fund}(\vec{a},\vec{Y},-\mu_4),
\label{eq:two_afund}
\end{eqnarray}
where $Z_{\rm fund}(\vec{a},\vec{Y},\mu)$ is a contribution from a fundamental hypermultiplet with mass $\mu$,
which is equivalent to our expression if  $\epsilon_+=0$  
\begin{eqnarray}
 Z_{\rm afund}(\vec{a},\vec{Y},\mu_3)Z_{\rm afund}(\vec{a},\vec{Y},\mu_4)
\end{eqnarray}
because $Z_{\rm afund}(\vec{a},\vec{Y},\mu) \equiv Z_{\rm fund}(\vec{a},\vec{Y},\epsilon_+-\mu)$.

From now on we concentrate on the case of $\vec{a}= (a,-a)$ of interest.
Defining the free energy of  instantons by
\begin{eqnarray}
 F^{\rm inst} \equiv (-\epsilon_1\epsilon_2)\log Z_{\rm inst}.
\end{eqnarray} 
we can expand $F^{\rm inst}$ in powers of $g$ (with the relation \eqref{eq:relation1})
\begin{eqnarray}
 F^{\rm inst}= F_0^{\rm inst} + \frac{\epsilon_+}{2}F_1^{\rm inst} + \mathcal{O}(g^2).
\end{eqnarray}
In addition, if we put $\mu_I=m$ as treated in the text, we have 
\begin{eqnarray}
 F_0^{\rm inst} &=& \frac{a^4+6 a^2 m^2+m^4}{2 a^2}q + \frac{13 a^8+100 a^6 m^2+22 a^4 m^4-12 a^2 m^6+5 m^8}{64 a^6}q^2
\nonumber\\[2mm]
&& + \frac{23 a^{12}+204 a^{10} m^2+51 a^8 m^4-48 a^6 m^6+45 a^4 m^8-28 a^2 m^{10}+9 m^{12}}{192 a^{10}}q^3
\nonumber\\[2mm]
&& + \frac{1}{32768 a^{14}}\left(2701 a^{16}+26440 a^{14} m^2+7164 a^{12} m^4-9000 a^{10} m^6\right.
\nonumber\\[2mm]
&&\qquad \left.+12190 a^8 m^8-13384 a^6 m^{10}+10908 a^4 m^{12}-5720 a^2 m^{14}+1469 m^{16}\right)q^4
\nonumber\\[2mm]
&& + \mathcal{O}(q^5),\nonumber
\\[4mm]
F_1^{\rm inst} &=& -\frac{2m \left(a^2+m^2\right)}{a^2}q -\frac{9 a^6m+11 a^4 m^3-9 a^2 m^5+5 m^7}{8a^6}q^2
\nonumber\\[2mm]
&&-\frac{38 a^{10}m+51 a^8 m^3-72 a^6 m^5+90 a^4 m^7-70 a^2 m^9+27 m^{11}}{48a^{10}}q^3
\nonumber\\
&&-\frac{1}{2048 a^{14}} \left(1257 a^{14}m+1791 a^{12} m^3-3375 a^{10} m^5+6095 a^8 m^7-8365 a^6 m^9\right.
\nonumber\\
&&\qquad\qquad\left. +8181 a^4 m^{11}-5005 a^2 m^{13}+1469 m^{15}\right)q^4\;+\;\mathcal{O}(q^5).
\label{eq:half_inst_NF=4}
\end{eqnarray}

\subsection{$N_F=3$ theory}

The instanton part is given by
\begin{eqnarray}
 Z_{\rm inst} &=& \sum_{\vec{Y}} \Lambda_3^{|\vec{Y}|} Z_{\rm vec}(\vec{a},\vec{Y})Z_{\rm afund}(\vec{a},\vec{Y},\mu_1)Z_{\rm afund}(\vec{a},\vec{Y},\mu_2)Z_{\rm afund}(\vec{a},\vec{Y},\mu_3),
\end{eqnarray}
where $\mu_1,\mu_2$ and $\mu_3$ are the masses of three hyper multiplets and $\Lambda_3$ is a dynamical scale. The Coulomb branch parameter $\vec{a}$ generally has two independent components. However, as we have seen in appendix \ref{app:AGT}, we only need to consider the case of $\vec{a}=(a,-a)$ in this paper. The  half-genus expansion of the free energy has the form
\begin{eqnarray}
 F_0^{\rm inst} &=& \frac{m}{2}\Lambda + \frac{a^2+m^2}{64 a^2}\Lambda^2 + \frac{a^4-6 a^2 m^2+5 m^4}{32768 a^6}\Lambda^4 + \frac{5 a^4 m^2-14 a^2 m^4+9 m^6}{1572864 a^{10}}\Lambda^6 + \cdots,
\nonumber\\
 F_1^{\rm inst} &=& 
-\frac{\Lambda}{2} -\frac{m}{32 a^2}\Lambda^2 +\frac{m \left(3 a^2-5 m^2\right)}{8192\, a^6}\Lambda^4 -\frac{m \left(5 a^4-28 a^2 m^2+27 m^4\right) }{786432\, a^{10}}\Lambda^6 + \cdots.
\nonumber\\
\label{eq:half_inst_NF=3}
\end{eqnarray}
where we set $\mu_1=\mu_2=0$ and $\mu_3=m$. 
Note that the first term in $F_0$ and $F_1$ is independent of the Coulomb branch parameter $a$.

\subsection{$N_F=2$ theory}

The instanton partition function  is given (with  $\vec{a}=(a,-a)$)
\begin{eqnarray}
 Z_{\rm inst} = \sum_{\vec{Y}} \Lambda_2^{2|\vec{Y}|}Z_{\rm vec}(\vec{a},\vec{Y})Z_{\rm afund}(\vec{a},\vec{Y},\mu_1)Z_{\rm afund}(\vec{a},\vec{Y},\mu_3).
\end{eqnarray}
The  half-genus expansion of the free energy has the form
\begin{eqnarray}
 F_0^{\rm inst} &=& \frac{a^2+m^2}{2a^2}\Lambda^2 + \frac{a^4-6 a^2 m^2+5 m^4}{64 a^6}\Lambda^4+\frac{5 a^4 m^2-14 a^2 m^4+9 m^6}{192\, a^{10}}\Lambda^6
\nonumber\\[3mm]
&& +\frac{5 a^8-252 a^6 m^2+1638 a^4 m^4-2860 a^2 m^6+1469 m^8}{32768\, a^{14}}\Lambda^8+\cdots,
\nonumber \\[5mm]
 F_1^{\rm inst} &=& -\frac{m}{a^2}\Lambda^2 +\frac{m \left(3 a^2-5 m^2\right) }{16 a^6}\Lambda^4-\frac{m \left(5 a^4-28 a^2 m^2+27 m^4\right)}{96\,a^{10}}\Lambda^6
\nonumber\\[3mm]
&& +\frac{m \left(63 a^6-819 a^4 m^2+2145 a^2 m^4-1469 m^6\right)}{4096\, a^{14}}\Lambda^8 + \cdots
\label{eq:half_inst_NF=2}
\end{eqnarray}
where we set $\mu_1=\mu_3=m$.

\end{document}